# Electromagnetic pulses which have a zero momentum frame

# John Lekner

School of Chemical and Physical Sciences, Victoria University of Wellington

P O Box 600, Wellington, New Zealand

One set of the Ziolkowski family of exact solutions of the wave equation are shown to represent pulses propagating with momentum smaller than energy/$c$. This is explicitly demonstrated for special cases by calculating the total electromagnetic momentum and energy. The ratio of momentum to energy is a constant smaller than $c^{-1}$ and so there exists a Lorentz transformation to a frame in which the total momentum is zero. In the zero-momentum frame the fields are those of an annular pulse converging onto or diverging from a focal region.

It is axiomatic in special relativity that there is no rest frame for light: the speed of light is the same in every inertial frame. It is known, however, that electromagnetic energy can travel at less than the speed of light [1], and here we shall show that there exists a Lorentz frame $L_0$ in which the total pulse electromagnetic momentum is zero, for a class of solutions of the Maxwell equations.

Ziolkowski [2] obtained exact solutions of the wave equation $\nabla^2 \psi = c^{-2} \partial^2 \psi / \partial t^2$ in the form ($\rho = \sqrt{x^2 + y^2}$ is the distance from the propagation axis)

$$\psi(\mathbf{r}, t) = \int_0^\infty dk\, F(k) \frac{\exp\{ik(z+ct) - k\rho^2/[b+i(z-ct)]\}}{b+i(z-ct)} \tag{1}$$

With $F(k) = abe^{-ka}$ one obtains the particularly simple solution [2, 3, 4]

$$\psi(\mathbf{r}, t) = \frac{ab}{\rho^2 + [a-i(z+ct)][b+i(z-ct)]} \psi_0 \tag{2}$$

where $\psi_0$ is the wavefunction value at the space-time origin. Feng, Winful and Hellwarth [4] have called the electromagnetic fields derived from (2) "focused single-cycle electromagnetic pulses". We shall show below that, for an arbitrary electromagnetic pulse, the total momentum $\mathbf{P}$ and total energy $U$ are both constant in time. We also show that the fields derived from (2) have $cP_z < U$. Thus a Lorentz transformation to a zero-momentum frame $L_0$ is possible, and in that frame the fields represent an annular pulse, converging for $t_0 < 0$, diverging for $t_0 > 0$. (In the 'lab' frame the wavefunction (2) gives fields in which there is net forward or backward propagation, in general.)

Given solutions of the wave equation, solutions of Maxwell's equations can be obtained as $\mathbf{E} = -\nabla\Phi - \partial_t \mathbf{A}, \mathbf{B} = \nabla \times \mathbf{A}$, where $\Phi$ and all components of $\mathbf{A}$ satisfy $\nabla^2 \psi = \partial_t^2 \psi$ (here $\partial_t$ denotes differentiation with respect to $ct$), provided the Lorentz condition $\nabla \cdot \mathbf{A} + \partial_t \Phi = 0$ holds [5]. For example, we can take

$\Phi = 0$, $\mathbf{A} = \nabla \times [0, 0, \psi] = [\partial_y, -\partial_x, 0]\psi$; this gives a TE field $\mathbf{E} = [-\partial_y \partial_t, \partial_x \partial_t, 0]\psi$,

$\mathbf{B} = [\partial_x \partial_z, \partial_y \partial_z, -\partial_x^2 - \partial_y^2]\psi$. A TM field is obtained by the duality transformation $\mathbf{E} \to \mathbf{B}$,

$\mathbf{B} \to -\mathbf{E}$: $\mathbf{E} = \nabla \times \mathbf{A}$, $\mathbf{B} = \partial_t \mathbf{A}$. The combination TE + iTM has $\mathbf{E} = -\partial_t \mathbf{A} + i\nabla \times \mathbf{A}$,

$\mathbf{B} = \nabla \times \mathbf{A} + i\partial_t \mathbf{A}$, ie $\mathbf{E} = i\mathbf{B}$ (TE – iTM gives $\mathbf{E} = -i\mathbf{B}$). In the monochromatic beam case these combinations give *steady beams*, in which the electromagnetic energy density $u$ and momentum density $\mathbf{p}$ (= Poynting vector $/c^2$) do not oscillate in time [6, 1]. As in the steady beam case, the $\mathbf{E} = \pm i\mathbf{B}$ solutions have (taking either Re($\mathbf{E}$, $\mathbf{B}$) or Im($\mathbf{E}$, $\mathbf{B}$) as the physical fields)

$$u = \frac{1}{8\pi}|\mathbf{E}|^2 = \frac{1}{8\pi}|\mathbf{B}|^2, \quad \mathbf{p} = \frac{i}{8\pi c}\mathbf{E} \times \mathbf{E}^* = \frac{i}{8\pi c}\mathbf{B} \times \mathbf{B}^* \qquad (3)$$

When $\psi(\mathbf{r}, t)$ is independent of the azimuthal angle $\phi$, and $\mathbf{A} = [\partial_y, -\partial_x, 0]\psi$, we find

$$u = \frac{1}{8\pi}\left\{|\partial_\rho \partial_z \psi|^2 + |\partial_\rho \partial_t \psi|^2 + |\partial_z^2 \psi - \partial_t^2 \psi|^2\right\}, \quad p_z = -\frac{1}{4\pi c}\text{Re}\{(\partial_\rho \partial_t \psi^*)(\partial_\rho \partial_z \psi)\} \qquad (4)$$

and $p_x = p_\rho \cos\phi - p_\phi \sin\phi$, $p_y = p_\rho \sin\phi + p_\phi \cos\phi$, where the radial and azimuthal components of the momentum density are given by

$$p_\rho = \frac{1}{4\pi c}\text{Re}\{(\partial_\rho \partial_t \psi^*)(\partial_z^2 \psi - \partial_t^2 \psi)\}, \quad p_\phi = \frac{1}{4\pi c}\text{Im}\{(\partial_\rho \partial_z \psi^*)(\partial_z^2 \psi - \partial_t^2 \psi)\} \qquad (5)$$

Figure 1 shows contours of $u$ and a field plot of $p_z$, $p_\rho$ for $a = b$ and $a = 2b$ at $ct = 0$ and $3b$.

It follows from Maxwell's equations that the total energy $U = \int d^3 r\, u(\mathbf{r}, t)$ is independent of time: $4\pi \partial_t U = \int d^3 r\, (\mathbf{E} \cdot \nabla \times \mathbf{B} - \mathbf{B} \cdot \nabla \times \mathbf{E}) = -\int d^3 r\, \nabla \cdot (\mathbf{E} \times \mathbf{B})$ (real fields) from the Maxwell curl equations, and this integral can be expressed as a surface integral at infinity, which is zero at finite times. Likewise, again with real fields and $\mathbf{P} = \int d^3 r\, \mathbf{p}(\mathbf{r}, t)$,

$4\pi \partial_t \mathbf{P} = -\int d^3 r\, [\mathbf{E} \times (\nabla \times \mathbf{E}) + \mathbf{B} \times (\nabla \times \mathbf{B})]$, and integrations by parts show that this is zero at finite times also. Thus the total energy and total momentum are constant in time, as we would expect. If the ratio $cP_z/U$ is less than unity, as we shall demonstrate it is in particular cases below, we can Lorentz-transform to the zero-momentum frame.

Since $U$ and $\mathbf{P}$ are independent of time, we can evaluate them at $t = 0$. For $\psi$ given by (2) and $A = [\partial_y \psi, -\partial_x \psi, 0]$ we have from (4) that, for the TE + iTM pulse,

$$u(\mathbf{r}, 0) = \frac{(ab\psi_0)^2}{\pi} \frac{r^4 + 2(a^2 + b^2 - ab)\rho^2 + (a^2 + b^2)z^2 + (ab)^2}{\left[r^4 + 2ab\rho^2 + (a^2 + b^2)z^2 + (ab)^2\right]^3} \tag{6}$$

$$cp_z(\mathbf{r}, 0) = \frac{(ab\psi_0)^2}{\pi} \frac{(a^2 - b^2)\rho^2}{\left[r^4 + 2ab\rho^2 + (a^2 + b^2)z^2 + (ab)^2\right]^3} \tag{7}$$

where $r^2 = \rho^2 + z^2$. The integrations in spherical polar coordinates $(r, \theta, \phi)$ are helped by the substitution $\cos\theta = \frac{r^2 + ab}{(a-b)r}\tan\chi$ ($a > b > 0$ is assumed). We find

$$U = \frac{\pi}{8}\frac{a+b}{ab}\psi_0^2, \quad cP_z = \frac{\pi}{8}\frac{a-b}{ab}\psi_0^2 \tag{8}$$

(The transverse components of momentum integrate to zero.) Thus $cP_z/U = (a-b)/(a+b)$ is less than unity: the net momentum of the electromagnetic field is less than its energy / c. We can interpret $c^2 P_z /U$ as an average energy velocity [5, 1] $\beta c$, $\beta = (a-b)/(a+b)$. Independently of this interpretation, the fact that $\beta < 1$ implies that we can transform to the Lorentz frame $L_0$ in which the total momentum is zero. In this frame we have $\rho$ unchanged, and

$$z = (z_0 + \beta ct_0)\big/\sqrt{1-\beta^2} \quad ct = (ct_0 + \beta z_0)\big/\sqrt{1-\beta^2} \tag{9}$$

and so, replacing $(1+\beta)/(1-\beta)$ by $a/b$,

$$z + ct = \sqrt{\frac{a}{b}}(z_0 + ct_0), \quad z - ct = \sqrt{\frac{b}{a}}(z_0 - ct_0) \tag{10}$$

The wavefunction in (2) thus becomes, in the zero-momentum frame,

$$\psi(\mathbf{r}_0, t_0) = \frac{ab\psi_0}{\rho^2 + \left[\sqrt{ab} - i(z_0 + ct_0)\right]\left[\sqrt{ab} + i(z_0 - ct_0)\right]} \tag{11}$$

which gives equal weight to the forward and backward propagations in the scalar wave.

Feng, Winful and Hellwarth [4] have taken vector potential $\mathbf{A} = \nabla \times [\psi, 0, 0]$, and fields $\mathbf{E} = -\partial_t \mathbf{A}$, $\mathbf{B} = \nabla \times \mathbf{A}$. This is a TE pulse, for which the calculations are more complicated than for the TE + iTM pulse above. For the TE pulse we find, for both the real and imaginary parts of $\psi$,

$$U = \frac{\pi}{64} \frac{(a+b)(3a^2 - 2ab + 3b^2)}{(ab)^2} \psi_0^2 \tag{12}$$

$$cP_z = \frac{\pi}{64} \frac{(a-b)(3a^2 + 2ab + 3b^2)}{(ab)^2} \psi_0^2 \tag{13}$$

(The expression for $U$ is in agreement with equation (3.6) of [4].) The Lorentz boost to $L_0$ is

$$\beta = \frac{a-b}{a+b} \frac{3a^2 + 2ab + 3b^2}{3a^2 - 2ab + 3b^2} \tag{14}$$

The wavefunction in the zero-momentum frame is now

$$\psi(\mathbf{r}_0, t_0) = \frac{ab\psi_0}{\rho^2 + \left[a/\alpha - i(z_0 + ct_0)\right]\left[b\alpha + i(z_0 - ct_0)\right]} \tag{15}$$

where

$$\alpha = \sqrt{\frac{1+\beta}{1-\beta}} = \sqrt{\frac{a(3a^2 + b^2)}{b(a^2 + 3b^2)}} \tag{16}$$

The angular momentum density is $\mathbf{j} = \mathbf{r} \times \mathbf{p}$ [5, 7] where $\mathbf{p}$ is the momentum density; the angular momentum of a pulse is $\mathbf{J} = \int d^3r\, \mathbf{j}$. For the TE + iTM and TE classical electromagnetic pulses described above, all the components of $\mathbf{J}$ are zero. By analogy with a 'steady' beam which in the plane-wave limit is circularly polarized everywhere [8], we construct the $\mathbf{E} = i\mathbf{B}$ pulse

$$\mathbf{A} = \nabla \times [-i\psi, \psi, 0], \quad \mathbf{E} = -\partial_t \mathbf{A} + i\nabla \times \mathbf{A}, \quad \mathbf{B} = \nabla \times \mathbf{A} + i\partial_t \mathbf{A} \tag{17}$$

The energy density and momentum density are again given by (3). The complex magnetic field is

$$\mathbf{B} = [(\partial_x + i\partial_y)\partial_y + i(\partial_z - \partial_t)\partial_z, \ -(\partial_x + i\partial_y)\partial_x - (\partial_z - \partial_t)\partial_z, \ -i(\partial_x + i\partial_y)(\partial_z - \partial_t)]\psi$$

(18)

and when $\psi$ is given by (2) we find the energy and z-components of momentum and angular momentum of the pulse to be

$$U = \frac{\pi}{8}\frac{3a+b}{b^2}\psi_0^2, \quad cP_z = \frac{\pi}{8}\frac{3a-b}{b^2}\psi_0^2, \quad cJ_z = -\frac{\pi}{4}\frac{a}{b}\psi_0^2 \quad (19)$$

The sign of $J_z$ is reversed if one takes the vector potential and fields to be

$$\mathbf{A} = \nabla \times [i\psi, \psi, 0], \quad \mathbf{E} = -\partial_t\mathbf{A} + i\nabla \times \mathbf{A}, \quad \mathbf{B} = \nabla \times \mathbf{A} + i\partial_t\mathbf{A} \quad (20)$$

The results are then as in (19) with $a$ and $b$ interchanged, and the sign of $J_z$ changed:

$$U = \frac{\pi}{8}\frac{a+3b}{a^2}\psi_0^2, \quad cP_z = \frac{\pi}{8}\frac{a-3b}{a^2}\psi_0^2, \quad cJ_z = \frac{\pi}{4}\frac{b}{a}\psi_0^2 \quad (21)$$

In general, since $P_z$ and $U$ have been shown to be independent of time, their ratio is also independent of time. Whenever $cP_z < U$, we can transform to a zero-momentum frame. We suspect that this is possible for all solutions which represent pulses converging onto and then diverging from a focal region.

According to Adlard, Pike and Sarkar [9], "given any classical solution of the source-free Maxwell's equations it is possible to write down a corresponding quantum mechanical one-photon state". They use a particular case of the Ziolkowski solutions to show that "single-photon states with arbitrarily high powers of asymptotic falloff can be explicitly constructed". The quantum fields resulting from the Ziolkowski family of solutions represent (for $t > 0$) photons diverging from a focal region. They are rather different from the textbook photon, which is monochromatic and unidirectional, with $U = \hbar\omega$ and $J_z = \pm\hbar$ or 0. Although there is no oscillation in the pulses derived from (2), we can associate an effective frequency with the pulse as follows: since ($c\mathbf{P}, U$) is a four-vector, the pulse energy in the zero-momentum frame $L_0$ is $U_0 = (U - \beta cP_z)/\sqrt{1-\beta^2}$. From (21), for example, we obtain (with $\beta = (a-3b)/(3a+b)$),

$$U_0 = \frac{\pi}{4}\frac{b}{a}\sqrt{\frac{3}{ab}}\,\psi_0^2 \qquad (22)$$

Thus, for the vector potential and fields in (20),

$$U = \frac{a+3b}{2\sqrt{3ab}}U_0, \quad cP_z = \frac{a-3b}{2\sqrt{3ab}}U_0, \quad cJ_z = \sqrt{\frac{ab}{3}}U_0 \qquad (23)$$

If we set $J_z = \hbar$ and $U_0 = \hbar\omega_0$, the resulting angular frequency in $L_0$ is

$$\omega_0 = U_0/J_z = c\sqrt{\frac{3}{ab}} \qquad (24)$$

The effective angular frequency in the lab frame is

$$\omega = U/J_z = c\frac{a+3b}{2ab} \qquad (25)$$

The frequency ratio $\omega/\omega_0$ is not given by the usual Doppler expression $\sqrt{\frac{1+\beta}{1-\beta}}$, since that applies only to monochromatic plane waves.

Stimulating conversations with Damien Martin are gratefully acknowledged.

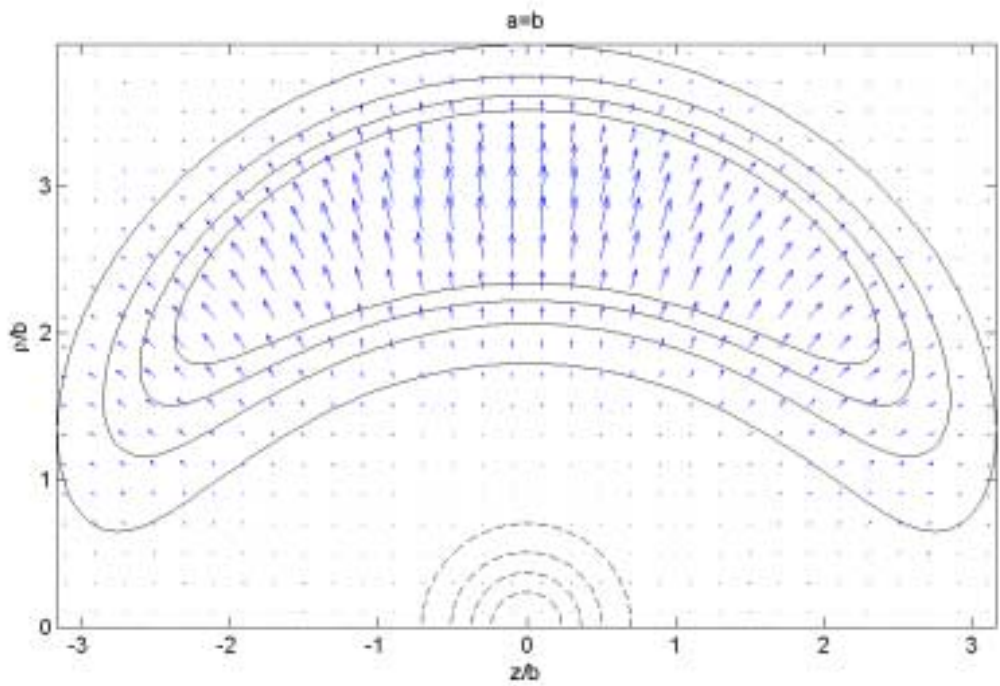

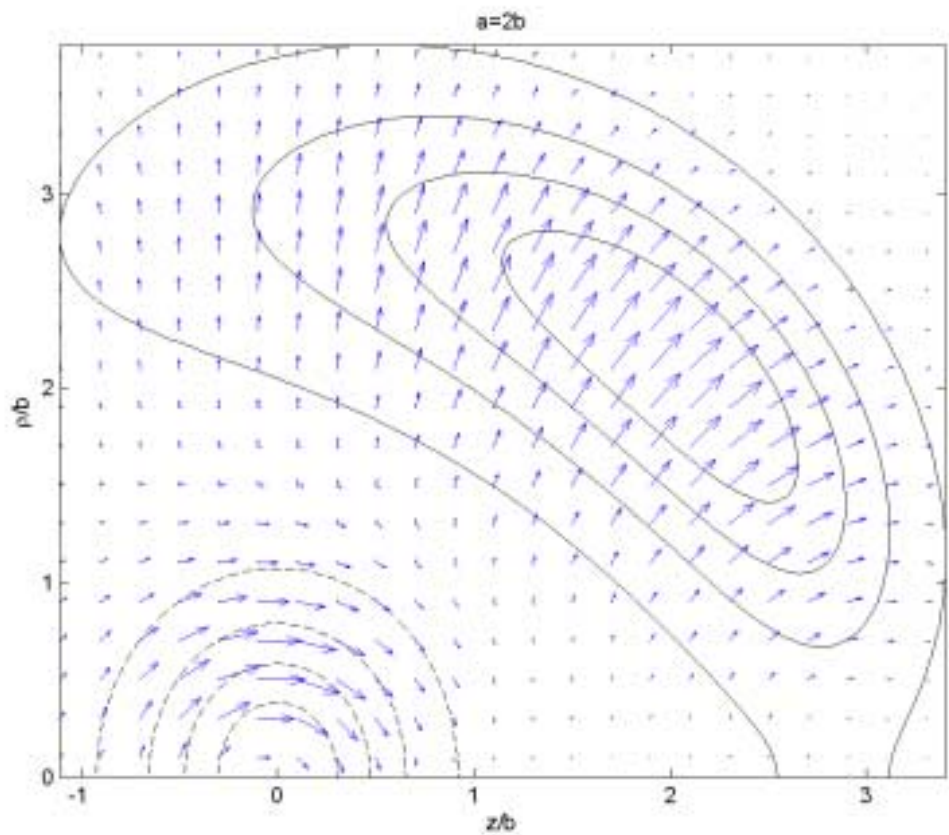

**Figures caption**

Both figures show contours of energy density for the TE + iTM pulse with $a = b$ (Figure 1) and $a = 2b$ (Figure 2). The contours are at 0.8, 0.6, 0.4 and 0.2 of the maximum at the given value of $ct$. In each case the $t = 0$ contours are dashed curves, and the $ct = 3b$ contours are solid curves. The three-dimensional contour surfaces are obtained by rotating the figures about the horizontal (z) axis. In the $a = b$ case the pulse is diverging (for $t > 0$) symmetrically from the origin, and the pulse momentum is zero. In the $a = 2b$ case there is a net momentum in the z direction, and the pulse energy density is asymptotically maximum on a cone of half-angle $\theta_m \approx 46.7°$. The general expression for the asymptotic angle at which the energy density is maximum is $\sin^2(\theta_m/2) = \left[5a - 3b - \sqrt{25a^2 - 46ab + 25b^2}\right]/8(a-b)$.

The arrows indicate the magnitude and direction of the projection of the momentum density onto a plane which includes the z-axis (ie the azimuthal component is not shown). The momentum density is identically zero at $t = 0$ when $a = b$. At $ct = 3b$ in the $a = 2b$ case, the magnitude of the momentum density has been increased by a factor of 22 (relative to the $t = 0$ values) for better visibility.